\documentclass[aps,pre,preprint,groupedaddress,showpacs]{revtex4-1}
\usepackage{graphicx}
\usepackage{amsmath}
\usepackage{amssymb}

\begin{document}
\title{Electric discharge contour dynamics model: the effects of curvature and finite conductivity.}
\author{M. Array\'as$^{1}$ and M. A. Fontelos$^{2}$}
\affiliation{$^{1}$\'Area de Electromagnetismo, Universidad Rey Juan
Carlos, Camino del Molino s/n, 28943 Fuenlabrada, Madrid, Spain}
\affiliation{$^{2}$Instituto de Ciencias Matem\'{a}ticas
(CSIC-UAM-UCM-UC3M), C/ Nicol\'{a}s Cabrera, 28049 Madrid, Spain}

\begin{abstract}
In this paper we present the complete derivation of the effective contour model for electrical discharges which appears as the asymptotic limit of the minimal streamer model for the propagation of electric discharges, when the electron diffusion is small. It consists of two integro-differential equations defined at the boundary of the plasma region: one for the motion and a second equation for the net charge density at the interface. We have computed explicit solutions with cylindrical symmetry and found the dispersion relation for small symmetry-breaking perturbations in the case of finite resistivity. We implement a numerical procedure to solve our model in general situations. As a result we compute the dispersion relation for the cylindrical case and compare it with the analytical predictions. Comparisons with experimental data for a 2-D positive streamers discharge are provided and predictions confirmed. 

\end{abstract}

\date{\today}
\maketitle

\section{Introduction}
The appearance and propagation of ionization waves is the prelude of
electrical breakdown of various media. In the case of a gas, the
specific features of the breakdown waves are determined by the type of
the gas, the value of the pressure, the geometry of the discharge cell
and the value and variation rate of the voltage at the electrodes. The
geometry determines the space distribution of the electric field and
hence the dynamics of the ionization fronts. In the case where there
in no initial ionization in the discharge gap, the ionization wave may
originate from one or several overlapping electron avalanches. After
attenuation of the electric field in the avalanche body, a conducting
channel or streamer develops: a plasma region fully ionized with a
positive side expanding towards the cathode and a negative region
towards the anode.
  
One of the approaches used to model the development of the
avalanche-streamer transition and the streamer propagation is a
nonlinear system of balance equations with a diffusion-drift
approximation for the currents, together with Poisson equation
\cite{Lagarkov}. Some progress in the understanding of the propagation
mechanism has been achieved using that model. We can mention: the
study of stationary plane ionization waves \cite{r1,vanS},
self-similar solutions for ionization waves in cylindrical and
spherical geometries \cite{aft,r3}, the effect of photoionization
\cite{mmt} and a branching mechanism as the result of the instability
of planar ionization fronts \cite{ME,aftprl,abft}. In this
hydrodynamic approximation, the fronts are subject to both stabilizing
forces due to diffusion which tend to dampen out any disturbances, and
destabilizing forces due to electric field which promote them. The
solution of the model, even in the simplest cases, poses a challenging
problem both numerical and analytical. Early numerical simulations can
be found in \cite{Dhali, Vitello}. Recently, a contour dynamics model
have been deduced in the limit of small electron diffusion
\cite{Arrayas10}, which resembles the Taylor-Melcher leaky dielectric
model for electrolyte solutions \cite{S}, but adapted to the context
of electric (plasma) discharges. This contour dynamics model allows to
study more general situations in two-dimensional and three-dimensional
cases.

\begin{figure}
\centering
\includegraphics[width=0.5\textwidth]{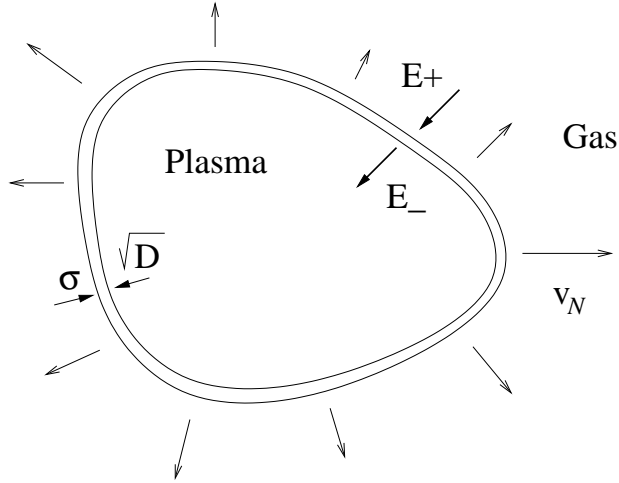}
\caption{The schematic of the contour dynamics model. The case displayed corresponds to a negative streamer discharge discharge so $\sigma$ represents the negative surface charge density. The electric field points towards the plasma region in this case.}
\label{fig1a}
\end{figure}

The contour dynamics model consists of an interface separating a plasma region from a neutral gas region as it is shown in Fig.~\ref{fig1a}. The separating surface has a net charge $\sigma$ and the thickness goes to zero as $\sqrt{D}$ being $D$ the charge diffusion coefficient. The case displayed in the figure correspond a negative discharge, so the electric field is pointing towards the plasma region and $\sigma$ is the negative charge density at the surface. The front will evolve following the equation
\begin{equation}
v_{N}=-\mu_e\mbox{E}_{\nu }^++2\sqrt{\frac{D_e}{l_0}\mu_e|\mbox{E}_{\nu}^+|\exp\left(-\frac{\mbox{E}_0}{|\mbox{E}_{\nu}^+|}\right)}-D_e\kappa, 
\label{cm1}
\end{equation}%
where $\mbox{E}_{\nu }^{+}$ is the normal component of the electric field at the
interface when approaching it from outside the plasma region, $\mu_e$ the eletron mobility, $D_e$ is the electron diffusion coefficient, E$_0$ is a characteristic ionization electric field and $\kappa$ the curvature of the interface. The parameter $l_0$ is the microscopic ionization characteristic length.  At the interface, the total negative surface charge density will change according to 
\begin{equation}
\frac{\partial \sigma }{\partial t}+\kappa v_{N}\sigma =-\frac{\mbox{E}_{\nu }^{-}}{\varrho_e}-j_\nu^{-}\,,
\label{cm2}
\end{equation}%
being now  $\mbox{E}_{\nu }^-$ the electric field at the interface coming from inside the plasma, $\varrho_e$ is a parameter proportional to the resistivity of the electrons in the created plasma and $j_\nu^{-}$ the current contribution of any electromotive force if present. 

Although the equations \eqref{cm1} and \eqref{cm2} are written for the
case of a negative front plotted in Fig.~\ref{fig1a}, and we will
present the derivation of the model for this case, we could use in
principle the same model for a positive front, but the electric field
should be sign reversed, and $\sigma$ would represent the positive
surface charge density. Although the moving carriers in the model are
the electrons, one may think of a front made of {\it holes}, with a
positive surface charge density, and characterized with the
corresponding parameters for the mobility, diffusion and so on.

In this paper, using the contour dynamics model, we will study
cylindrical discharges when the plasma has finite
conductivity. The dispersion curve for transversal instabilities will
be obtained for these finite conductivity streamers. The results will
be compared with the limiting cases of perfect conductivity, which is
the Lozansky-Firsov model \cite{LF} with a correction due to electron
diffusion, and with the case of a perfect insulator,
i.e the limit of very small conductivity. Finally, we compare the results with an actual experiment for a positive streamer discharge.  

We start by introducing the model. Taking a minimal set of balance equations to describe in fully a deterministic manner the discharge (see for example \cite{ME}), we will derive the contour dynamics equations for the evolution of the interface between the plasma region and the gas region free of charge (or with a very small density of charge). The outline of this derivation has already been reported \cite{Arrayas10} but here we present it in full details. Then, we proceed by studying a cylindrical discharge in the case of finite conductivity, and the analytical limits of infinite resistivity and ideal conductivity. With the model at hand we will predict some features of the stability of the fronts. Numerical simulations are made to calculate the dispersion curves and test some of the analytical predictions. We briefly describe the numerical methods employed in the corresponding section. We end with an analysis of the results, the comparison with an experiment for a positive 2-D streamer discharge, and overview of possibilities that the model opens for more complicated geometries and fully 3D cases.   
  
\section{The dynamical contour model}
In this section we obtain our model as a limit of a set of balance equations describing a streamer discharge. We will first recall the minimal description of a streamer discharge and some of the properties of the traveling planar fronts, and then make use of the asymptotic behaviour of those planar fronts in the limit of small diffusion to give a correction to the velocity of propagation of curved fronts. After finding the dynamics of the effective interface, a balance of the charge transport along the interface will be provided in order to complete the model.   

\subsection{The minimal model} 
For simulating the dynamical streamer development of streamers out of
a macroscopic initial ionization seed, in a non-attaching gas like
argon or nitrogen, the model of a streamer discharge \cite{contphys}
can be simplified. As a first approach, the processes with the smaller probabilities or cross sections can be ignored. Attachment and recombination processes can be neglected on that basis in comparison with the ionization process for non-attaching gases. We also ignore photoionization processes in this work. With these considerations in mind, the resulting balance equations are
\begin{eqnarray}
\frac{\partial N_{e}}{\partial t} &=& \nabla \cdot \left( \mu_e N_e \mbox{\bf E} + D_e\nabla N_e \right) + \nu_{i} N_e,
\label{model01} \\
\frac{\partial N_{p}}{\partial t} &=& \nu_{i} N_e, \label{model02}
\end{eqnarray}
where $N_{e}$ is the electron density, $N_{p}$ is the positive ion density, $\mu_e$ is the electron mobility and $D_e$ the diffusion coefficient. The ionization coefficient $\nu_{i}$ can be modeled following the phenomenological approximation suggested by Townsend, which leads to 
\begin{equation} 
\nu_{i} = \mu_e l_0^{-1}|\mbox{\bf E}|\exp \left(- \frac{\mbox{E}_0}{|\mbox{\bf E}|} \right),
\label{townsend}
\end{equation}
where  $l_{0}$ is the ionization length, and $\mbox{E}_{0}$ is
the characteristic impact ionization electric field. The fitting of experimental data can be done using those parameters \cite{Rai}. Note also that it is assumed the positive ions do not move and $\mu_e{\mbox{\bf E}}$ is the drift velocity of electrons. Those are valid approximations at the initial stages of the streamers development, but it may not be right afterward. To close the model, we consider Gauss's law
\begin{equation}
 \nabla \cdot {\mbox{\bf E}} = \frac{e(N_p-N_e)}{\varepsilon_0}.
\label{gauss}
\end{equation}

For convenience the equations are reduced to dimensionless
form. Townsend approximation provides physical scales and intrinsic
parameters of the model if only impact ionization is present in the
gas \cite{vanS}. The units are given by the ionization length $l_0$,
the characteristic impact ionization field $\mbox{E}_0$, and the
electron mobility $\mu_e$. The velocity scale yields $U_0=\mu_e
\mbox{E}_0$, and the time scale $\tau_0=l_0/U_0$. Typical
values of these quantities for nitrogen at normal conditions are
$l_0\approx 2.3\,\mu\mathrm m$, $\mbox{E}_0 \approx 200$ kV/m, and
$\mu_e \approx 380\,\mathrm {cm^2/Vs}$. We introduce the dimensionless
variables ${\bf r}_d={\bf r}/l_0$, $t_d=t/\tau_0$, the dimensionless
field ${\bf E}_d={\mbox{\bf E}}/\mbox{E}_0$, the dimensionless
electron and positive ion densities $n_e=N_e/N_0$ and
$n_p=N_p/N_0$ with $N_0=\varepsilon_0 {\cal E}_0/(e l_0)$, and the
dimensionless diffusion constant $D=D_e/(l_0 U_0)$. From now on, all the quantites will be dimensionless unless othewise stated. Note however that we will not write the subindex $d$. Just for reference, the dimensionless model reads 
\begin{eqnarray}
\label{1}
\frac{\partial n_e}{\partial t} &=& \nabla\cdot (n_e {\bf E} + D\; \nabla n_e) + n_e \alpha(|{\bf E}|),\\
\label{2}
\frac{\partial n_p}{\partial t} &=& n_e \alpha(|{\bf E}|), \\
\label{3}
\nabla\cdot{\bf E}&=&n_p - n_e, \\
\label{ft}
\alpha(|{\bf E}|) &=& |{\bf E}| \exp(-1/|{\bf E}|),
\end{eqnarray}

\subsection{Planar fronts and boundary layer}
Using the minimal streamer model, we can compute traveling wave solutions in the planar case. We will assume that the plasma region is on the left and the front is moving toward the right. The traveling waves are solutions such as $n_e$ and $n_p$ decay exponentially at infinity. This means that we can take 
\begin{eqnarray}
n_e &=& A e^{-\lambda(x-vt)},\nonumber\\
n_p &=& B e^{-\lambda(x-vt)},\nonumber \\
{\bf E}&=& (\mbox{E}^+ + Ce^{-\lambda(x-vt)})\,\hat{\bf x}\nonumber,
\end{eqnarray}
asymptotically far ahead for the planar wave in the $\hat{\bf x}$ direction, being $\mbox{E}^+$ the value of the electric field at the infinity. Introducing these expressions into the minimal model equations we get the relation 
\begin{equation}
D\lambda^2 - (\mbox{E}^+ +v)\lambda + \alpha(|\mbox{E}^+|)=0,
\label{lambda}
\end{equation}
which has real solutions if and only if
\begin{equation}
v \ge - \mbox{E}^+ + 2 \sqrt{D \alpha(|\mbox{E}^+|)}.
\end{equation}
All initial data decaying at infinity faster than $A e^{-\lambda^*x}$, with $\lambda^*=1/\sqrt{D \alpha(|\mbox{E}^+|)}$, will develop traveling waves with velocity $v^* =- \mbox{E}^+ + 2 \sqrt{D \alpha(|\mbox{E}^+|)}$. Clearly, from the assumption that the plasma state is on the left, negative velocity solutions are unphysical. So in the case of a negative front, when E$^+$ is negative, the front will move at least with the drift velocity in the case that $D=0$. For positive fronts, the motion will be possible only if the creation of charge, given by the Townsend factor, and its diffusion can compensate the drift. A detailed discussion about the propagation mechanism can be found at \cite{vanS}.   

\begin{figure}
  \centering \includegraphics[width=0.45\textwidth]{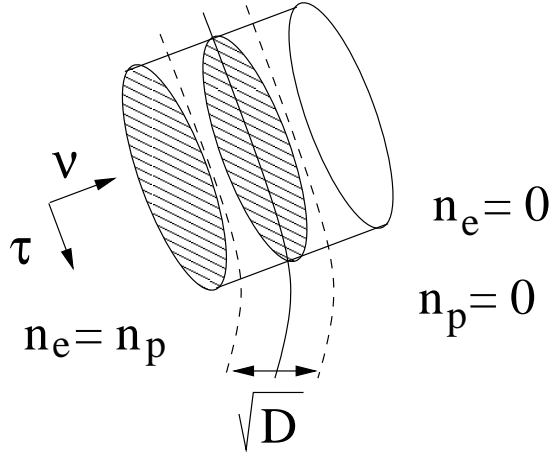}
\caption{Derivation of the contour dynamics model. We take a surface of constant $n_e$ at the boundary which has an effective width of order $\sqrt{D}$. The local coordinates tangent and normal to the surface, $\tau$ and $\nu$, together with a pillbox are also shown schematically.}
\label{fig2a}
\end{figure}

If $D\ll1$ the profiles for $n_p$ and ${\bf E}$ will vary very little from the profiles with $D=0$ and $n_e$ will develop a boundary layer at the front. This boundary layer has a width of $O(\sqrt{D})$ as shown in Fig.~\ref{fig2a}. The main results for the structure of the boundary layer which we are going to make use are
\begin{eqnarray}
\label{nebl}
n_e &=& f(\chi),\\
\label{npbl}
n_p &=& -\sqrt{D}\int_{\chi}^{\infty}f(z)\,dz,\\
\label{Ebl}
\mbox{E}&=&\mbox{E}^++O(\sqrt{D}),
\end{eqnarray} 
with $\chi = (x-v^*t)/\sqrt{D}$, and $\mbox{E}$ the electric field in the $\hat{\bf x}$ direction. The function $f(\chi)$, also appearing in \eqref{npbl}, is the solution of the equation
\begin{equation}
\frac{\partial ^{2}f}{\partial \chi ^{2}}+2\sqrt{\alpha(|\mbox{E}^+|)}\frac{\partial f}{\partial \chi} = f(f-1),
\label{blf} 
\end{equation}
which becomes the solution of a Fisher equation under the additional assumptions that the Townsend factor $\alpha(|{\bf E}|) \approx 1$. So, as it is plotted in Fig.~\ref{fig2a}, the function $f$ changes from constant values in a region of width $\sqrt{D}$, when imposing the two maching conditions $f(-\infty)=1$ and $f(\infty)=0$, thus separating the plasma region from the gas. The complete mathematical details can be found in \cite{aftprl} and \cite{abft}.     

\subsection{The correction due to the curvature}
Next we will add the correction to the propagation velocity due to the curvature of the front.  We take a level surface of $n_e$ representing the interface, and
introduce local coordinates $\tau $ (along the level surfaces of $n_{e}$) and $\nu$ (orthogonal to the level surfaces of $n_{e}$). The schematic can be seen in Fig.~\ref{fig2a}. We scale the normal coordinate with the boundary layer thickness $\nu=\chi\sqrt{D}$, and expand the Laplacian times $D$ like 
\[
D\,\Delta =\frac{\partial ^{2}}{\partial \chi ^{2}}+\sqrt{D}\kappa \frac{%
\partial }{\partial \chi }+D\left( \Delta _{\perp}-\kappa ^{2}\chi \frac{%
\partial }{\partial \chi }\right) +O(D^{\frac{3}{2}}),
\]%
where $\Delta\equiv\nabla^2$ is the Laplacian operator, $\Delta _{\perp}$ is the transverse Laplacian and $\kappa$ is twice the mean curvature in 3-D or just the curvature in 2-D (details of this expansion can be found in \cite{Pismen}). We write \eqref{1} in local coordinates, and using \eqref{3}, we find
\[\begin{split}
\frac{\partial n_{e}}{\partial t}-\mbox{E}_{\tau }\frac{\partial n_{e}}{\partial
\tau }-\left( \frac{\mbox{E}_{\nu }}{\sqrt{D}}+\sqrt{D}\kappa \right) \frac{%
\partial n_{e}}{\partial \chi }=\\ =\frac{\partial ^{2}n_{e}}{\partial \chi ^{2}}%
+n_{e}\alpha(|\mathbf{E}|)+n_e\left(n_{p}-n_{e}\right) +O(D).
\end{split}
\]%
Finally we use \eqref{blf} so that
\begin{equation}
\frac{\partial n_{e}}{\partial t}-\mbox{E}_{\tau }\frac{\partial n_{e}}{\partial
\tau }-\left( \frac{\mbox{E}_{\nu }}{\sqrt{D}}-2\sqrt{\alpha(|\mbox{E}_{\nu }|)}+\sqrt{D}\kappa \right) \frac{%
\partial n_{e}}{\partial \chi }=O(D^{\frac{1}{2}}).  \label{transne}
\end{equation}%
Note that the curvature term correction will be relevant provided $ 1\ll \kappa \ll D^{-\frac{1}{2}}$. Thus we have obtained a transport equation for the electron density
with velocity
\begin{equation}
\mathbf{v}=-\mathbf{E}+(2\sqrt{D\alpha(|\mathbf{E}|)}-D\kappa )\mathbf{n},
\label{vkurvature} 
\end{equation}
The level line $n_e$ which we have taken as representative of the interface evolution will move with a normal velocity
\begin{equation}
v_{N}=-\mbox{E}_{\nu }+2\sqrt{D\alpha(|\mathbf{E}|)}-D\kappa.  \label{bb1}
\end{equation}%
Notice that the level lines concentrate in a small region where $n_e$
presents a jump from its bulk value to zero, so most level lines
follow \eqref{bb1}. The tangential component of the velocity will not
change the geometry of the interface during its evolution, although
tangential exchanges of charge affect the evolution through the
dependence of $v_{N}$ on $E_{\nu }$. The mathematical description of
this effect will be the subject of next section.

\subsection{Charge transport along the interface}
In order to describe the charge transport along the interface we trace a small ``pillbox'' $\mathcal{D}$ around a portion of the interface having the top and bottom areas bigger than the lateral area, i.e. $\Delta \tau \gg \Delta \nu$ as we can see in Fig.~\ref{fig2a}. On the other hand $\mathcal{D}$ will be big enough to contain the diffusive layer and so the portion where the total negative charge density $n_{e}-n_{p}$ has significant values different of zero. 

We subtract \eqref{1} from \eqref{2} and integrate over the pillbox volume $\mathcal{D} $, assume that $n_{e}\rightarrow 0$ for $\chi =\nu /\sqrt{D}\gg 1$,
$\left\vert \nabla n_{e}\right\vert \rightarrow 0$ for $\left\vert \chi
\right\vert \gg 1$ and get
\begin{equation}
\frac{\partial}{\partial t}\int_{\mathcal{D} }(n_{e}-n_{p})\,dV = \left. n_{e}\mbox{E}_{\nu }\Delta \tau \right|_{\chi = -\infty }^{\infty }+O(D^{\frac{1}{2}}),
\label{s1}
\end{equation}
 where the contributions of the lateral transport of charge through the lateral surface
 is neglected in comparison with the exchange of charge in the normal direction. Note that in the Taylor-Melcher model this assumption is also made. As explained in \cite{Arrayas10} the left hand side of equation \eqref{s1} can be written as the time partial derivative of the product of the negative surface charge density $\sigma$ times the normal area $\Delta\tau$, and the change of a surface element can be related to the curvature times the normal velocity, so that
\begin{equation}
\frac{\partial \sigma }{\partial t}+\kappa v_{N}\sigma= -\left. n_{e}\mbox{E}_{\nu }\right|_{\chi = -\infty }
\label{s2}
\end{equation}

If a charge source $I(t)$ is present in the plasma, for instance at $x_0$, this source will create a current density inside the plasma and we will have at the interior of $\Omega$
\begin{equation}
\nabla \cdot \mathbf{j} = I(t)\delta(\mathbf{x}-\mathbf{x}_0).
\label{nablaj}
\end{equation}
By adding this contribution to \eqref{s2} we can finally write
\begin{equation}
\frac{\partial \sigma }{\partial t}+\kappa v_{N}\sigma =-\frac{\mbox{E}_{\nu }^{-}}{\varrho}-j_\nu^{-}\,,
\label{sigmaeq}
\end{equation}%
where $j_\nu^{-}$ is the current density coming from the ionized region $\Omega$ to its boundary $\partial \Omega$ in the normal direction $\nu$, $\mbox{E}_{\nu }^{-}$ is the normal component of the electric field when approaching the interface from inside, and $\varrho^{-1}=\lim_{\chi =-\infty }n_{e}$ is the effective movility of the electrons inside the plasma. Note that the quasineutrality of the plasma, further away of the interface is not changed by the current, but there is a jump in the normal component of the electric field across
the interface given by 
\begin{equation}
\mbox{E}_{\nu }^{+}-\mbox{E}_{\nu }^{-}=-\sigma, 
\label{jump}
\end{equation}%
with $\mbox{E}_{\nu }^{+}$ the normal component of the electric field when approaching the interface from outside the plasma region. 

\subsection{The effective contour model}
The Eqs.~\eqref{bb1} and \eqref{sigmaeq} together constitute the dynamical model able to describe the evolution of an interface separating a plasma region from a neutral region. Notice that in the case $\varrho^{-1}\gg 1$, we arrive to Lozansky-Firsov
model \cite{LF} with a correction due to electron diffusion, meanwhile in
the limit $D = 0$ we arrive at the classical Hele-Shaw model. Such a
model is known to possess solutions that develop singularities in the
form of cusps in finite time \cite{PK} but, when regularized by
surface tension corrections, the interface
may develop various patterns including some of fractal-type (see
\cite{Low} for a recent development and references therein).

Eq.~(\ref{sigmaeq}) will provide the surface charge density $\sigma $
as a function of time. From it, we can compute the electric field and move the interface with (\ref{bb1}). Two limits can be easily identified in the case that there is no charge injection inside the plasma, i.e $\mbox{j}_{\nu}^-\approx 0$: a) the limit of large conductivity%
\[
\varrho^{-1}\gg 1, \ \mbox{E}_{\nu }^{-}=0,
\]%
so that the interface is equipotential and b) the limit of small
conductivity%
\[
\varrho^{-1}\ll 1,\ \frac{\partial }{\partial t}\left( \sigma \Delta \tau \right)
=0\Rightarrow \sigma \Delta \tau =Cte,
\]%
where the charge contained by a surface element is constant and the density only changes through deformation (with change of area) of the interface. In the next sections we will study the intermediate case of finite resistivity. 

\section{The case of finite resistivity in 2-D geometries}
As an application we will solve the 2-D case for different conductivities. In order to grasp some features of the model first we will consider how fronts with radial symmetry evolve. Then we will study the stability of those fronts under small perturbations and finally solve the model numerically in order to test some of the analytical predictions.  

\subsection{Solutions with radial symmetry}
The electric potential created by a surface charge distribution with radial symmetry at the distance $r$ is found by solving the equation
\begin{equation}
\Delta V=\sigma \delta(r). 
\end{equation}
The fundamental solution turns out to be in polar coordinates
\begin{equation}
V(\mathbf{x})=\begin{cases}
    C\log |\mathbf{x}|,&  |\mathbf{x}| > r\\
    C\log r, & |\mathbf{x}| \le r
  \end{cases}
\end{equation}
where $C$ will be determined by the condition of the electric field jump \eqref{jump} at the surface. From the potential solution we can compute the electric field which has a discontinuity at the surface
\begin{equation}
E_{\nu }^{-}=0, \qquad  E_{\nu }^{+}=-\frac{C}{r}.
\end{equation}
For the current density, the solution of \eqref{nablaj} gives 
\begin{equation}
\mathbf{j}=\frac{I(t)}{2\pi r^2}\mathbf{r},
\end{equation} 
and finally using \eqref{sigmaeq} and the fact that $v_N = dr/dt$ and $\kappa = 1/r$, we get
\begin{equation}
\frac{\partial \sigma }{\partial t}+\frac{1}{r}\frac{\partial r}{\partial t}\sigma = -\frac{I(t)}{2\pi r}.
\label{sigma}
\end{equation}
This equation can be easily solved. We can write it as
\begin{equation}
\frac{\partial (r\sigma) }{\partial t} = -\frac{I(t)}{2\pi},
\end{equation} 
to get
\begin{equation}
\sigma = -\frac{Q(t)}{2\pi r}, \quad \text{with} \quad Q(t)= \int_0^t I(t)\,dt,
\end{equation}
where we have assumed that $\sigma(0)=0$. Now we can see from the condition \eqref{jump} that $C = -Q(t)/2\pi$, so 
\begin{equation}
E_{\nu}^{+}=\frac{Q(t)}{2\pi r}.
\end{equation}
Then, defining $\varepsilon\equiv D$, the interface evolves according to \eqref{cm1} as
\begin{equation}
\frac{dr}{dt}=-\left(\frac{Q(t)}{2\pi}+\varepsilon \right)\frac{1}{r}+2\sqrt{\varepsilon\alpha(|Q(t)/2\pi r|)}.
\label{r0}
\end{equation}

We shall analyze next two limiting cases. First the case where  
\begin{equation}
r \ll \frac{|Q(t)|}{4\pi \varepsilon^{\frac{1}{2}}\sqrt{\alpha(|Q(t)/2\pi r|)}}, \quad \text{and} \quad \varepsilon \ll 1.
\end{equation}
Then expression \eqref{r0} results
\begin{equation}
\frac{dr}{dt}\approx-\frac{Q(t)}{2\pi r},
\end{equation}
so 
\begin{equation}
 r(t)\approx \sqrt{r(0)^2 - \int_0^t Q(t')/\pi \,dt'}. 
\end{equation}
For the particular case $Q(t)=Q$ is constant
\begin{equation}
r(t)\approx \sqrt{r(0)^2 - t Q/\pi}
\end{equation}
The second case is the opposite one. If 
\begin{equation}
  r \gg \frac{|Q(t)|}{4\pi \varepsilon^{\frac{1}{2}}\sqrt{\alpha(|Q(t)/2\pi r|)}}, \quad \text{and} \quad \varepsilon \ll 1, 
\end{equation}
we have now 
\begin{equation}
\frac{dr}{dt}\approx 2\varepsilon^{\frac{1}{2}}\sqrt{\alpha(|Q(t)/2\pi r|)}.
\end{equation}
For the particular case $Q(t)= Q$, by standard asymptotic calculations, when $t\gg 1$ we deduce
\begin{equation}
\quad r(t)\approx \frac{|Q|}{\pi}\log{t}.
\end{equation}

\subsection{Stability analysis}
We will study now the stability of the fronts under small perturbations. 
We change by a small amount the position of the front as well as the charge density. 
The perturbed position and charge surface density of the interface on the interface will be parametrized using the polar angle as
\begin{eqnarray}
  \label{pert1}
  r(\theta,t)=r(t)+\delta S(\theta,t),\\
  \label{pert2}
  \sigma(\theta,t)=-\frac{Q(t)}{2\pi r(\theta,t)}+\delta \Sigma(\theta,t),
\end{eqnarray}
where $r(t)$ is the solution of the equations for the radial symmetrical front, $Q(t)= \int_0^t I(t)\,dt$ and $\delta$ a small parameter.

The electric potential will change by $\delta V_p(\mathbf{x})$ after adding a geometrical perturbation of the interface and some extra charge on it. This term satisfies the equation $\Delta V_p = O(\delta)$. Changing coordinates to 
\[
\mathbf{x} \longrightarrow \mathbf{\tilde{x}}=\mathbf{x}\,\frac{r(t)}{r(\theta,t)},
\]
the perturbed surface becomes a disk of radius $r(t)$ again, and solving for it yieds
\begin{eqnarray}
V_p(\tilde{r},\theta)=\sum_1^{\infty}\psi_n \cos(n\theta)\left(\frac{r}{\tilde{r}}\right)^n,  \,\,\tilde{r} >  r\label{cvv1}\\
V_p(\tilde{r},\theta)=\sum_1^{\infty}\varphi_n \cos(n\theta)\left(\frac{\tilde{r}}{r}\right)^n,  \,\,\tilde{r} \le r
\label{cv1}
\end{eqnarray}
where it is imposed that $V_p$ remains finite at the origin and at very large distances becomes zero.

Now taking the condition of continuity for the potential, we have at the interface $\mathbf{x}_s$ (in the original coordinate system)
\[
V_p(\mathbf{x}_s^+)= V_p(\mathbf{x}_s^-) + S\,\frac{Q(t)}{2\pi r(t)},
\]
and writing the surface perturbation as
\begin{equation}
S = \sum_{n=1}^{\infty}s_n(t)\cos(n\theta),
\label{sn}
\end{equation}
the coefficients of the series in \eqref{cvv1} and \eqref{cv1} can be related by
\begin{equation}
\psi_n = \varphi_n + \frac{Q(t)}{2\pi r} s_n.
\label{coeff}
\end{equation}

Making use of the expressions \eqref{cvv1}--\eqref{coeff}, one can calculate the electric field to $\delta$ order. We will need the normal components of the electric field at both sides of the surface, together with the jump condition \eqref{jump} to find the charge perturbation of\eqref{pert2}. The normal components of the electric field at the interface are
\begin{eqnarray}
E_{\nu}^+&=&\frac{Q(t)}{2\pi (r+\delta S)}+\delta\sum_1^{\infty}\left(\varphi_n + \frac{Q(t)}{2\pi r}s_n\right)\frac{n}{r}\cos(n\theta),\nonumber \\
E_{\nu}^-&=&-\delta\sum_1^{\infty}\varphi_n \frac{n}{r}\cos(n\theta),
\label{Enormal}
\end{eqnarray}
thus
\begin{equation}
\Sigma = -\sum_{n=1}^\infty \left(2\varphi_n + \frac{Q(t)}{2\pi r} s_n \right)\frac{n}{r}\cos(n\theta).
\label{sigman}
\end{equation}

The dynamics of the front will be changed by the perturbation introduced. The curvature correction turns out to be
\begin{equation}
\kappa = \frac{r^2+2rS\delta-rS_{\theta\theta}\delta+O(\delta^2)}{\left(r^2+2rS_{\theta}\delta+O(\delta^2)\right)^{\frac{3}{2}}} = \frac{1}{r}-\frac{S+S_{\theta\theta}}{r^2}\delta+O(\delta^2),
\label{kappap}
\end{equation}
(the subindex $\theta$ means the partial derivative with respect this variable) and the normal component of the velocity 
\begin{equation}
v_N = \frac{d r(t)}{d t} +\delta \frac{\partial S(\theta,t)}{\partial t},
\label{vp}
\end{equation}
so the contour model equation \eqref{bb1}, to first order gives
\begin{widetext}
\begin{equation}
\begin{split}
\frac{d r(t)}{d t} + \delta\frac{\partial S(\theta,t)}{\partial t} 
&= -\frac{Q(t)}{2\pi r}+\delta S\frac{Q(t)}{2\pi r^2}  -\delta\sum_1^{\infty}\left(\varphi_n + \frac{Q(t)}{2\pi r}s_n\right)\frac{n}{r}\cos(n\theta) +  2\varepsilon ^{\frac{1}{2}}\sqrt{\alpha_0+\delta \alpha_1}-\\&-\varepsilon  \left(\frac{1}{r}-\delta\,\frac{S+S_{\theta\theta}}{r^2}\right).  
\label{cmp1}
\end{split}
\end{equation} 
where we have written the Townsend function \eqref{ft} up to first order as $\alpha = \alpha_0 +\delta\alpha_1+ O(\delta^2)$. Now, we have
\[|E_0+\delta E_1|e^{-\frac{1}{|E_0+E_1\delta|}} \approx|E_0|e^{-\frac{1}{|E_0|}}+\delta\,\mathrm{sign}(E_0) E_1\biggl(1+\frac{1}{|E_0|}\biggr)e^{-\frac{1}{|E_0|}}=\alpha_0 +\delta\alpha_1,\]
where, using \eqref{Enormal},
\begin{eqnarray}
E_0&=&\frac{Q(t)}{2\pi r},\label{elec0}\\
E_1&=&\sum_{n=1}^{\infty}\left(n\varphi_n + (n-1)\frac{Q(t)}{2\pi r}s_n\right)\frac{1}{r}\cos(n\theta),
\end{eqnarray}
so that
\[
\sqrt{\alpha}=\sqrt{\alpha_0}+\delta\frac{\alpha_1}{2\sqrt{\alpha_0}}=\sqrt{\alpha_0}\biggl[1+\delta\,\mathrm{sign}(Q(t))\frac{E_1}{2|E_0|}\biggl(1+\frac{1}{|E_0|}\biggr)\biggr].
\]

Taking into account \eqref{r0} for the zero order term, we get from \eqref{cmp1}
\begin{equation}
\frac{\partial S}{\partial t} = S\frac{Q(t)}{2\pi r^2} -\sum_1^{\infty}\left(\varphi_n  + \frac{Q(t)}{2\pi r}s_n\right)\frac{n}{r}\cos(n\theta) + \varepsilon  \left(\frac{S+S_{\theta\theta}}{r^2}\right)+\varepsilon^\frac{1}{2}\frac{\alpha_1}{\sqrt{\alpha_0}},
\label{eqS}
\end{equation}
and finally making use of the expansion \eqref{sn} for the perturbation $S$ yields
\begin{equation}
\begin{split}
\label{eqSn}
 \frac{ds_n}{dt}& = \Biggl[-1+\varepsilon^\frac{1}{2}\frac{2\pi r\sqrt{\alpha_0}\,\mathrm{sign}(Q(t))}{|Q(t)|}\biggl(1+\frac{2\pi r}{|Q(t)|}\biggr)\Biggr]\frac{n}{r}\varphi_n-\\&-\Biggl[\frac{Q(t)}{2\pi r^2}(n-1)+\frac{\varepsilon}{r^2}(n^2-1)+\varepsilon^\frac{1}{2}\frac{(n-1)\sqrt{\alpha_0}}{r}\left(1+\frac{2\pi r}{|Q(t)|}\right)\Biggr]s_n.
\end{split}
\end{equation}
\end{widetext}

In order to find the correction to the charge density we take Eq.\eqref{sigmaeq} and
 multiply it by $r(\theta,t)$. Then we use the curvature expansion \eqref{kappap} written as
\[
\kappa =\frac{1}{r(\theta,t)}-\frac{S_{\theta \theta}}{r^2}\delta, 
\]
(being $r=r(t)$ is the zero order term in the position),
and the fact that \[v_N=\frac{dr(\theta,t)}{dt}.\]
Hence
\begin{widetext}
\[\frac{\partial (r(\theta,t) \sigma(\theta,t)) }{\partial t} - r(\theta,t)\frac{S_{\theta \theta}}{r^2}v_N \sigma(\theta,t)\,\delta = -\frac{r(\theta,t)}{\varrho}E_\nu^--\frac{I(t)}{2\pi},\]
so that, at $O(\delta)$,
\begin{equation}
\frac{\partial (r \Sigma)}{\partial t} + \frac{S_{\theta \theta}}{r}\frac{Q(t)}{2\pi r}\frac{dr}{dt}= \frac{1}{\varrho}\sum_1^{\infty}n\varphi_n \cos(n\theta). 
\label{eqSig}
\end {equation}
Making use of the \eqref{r0}, \eqref{sn} and \eqref{sigman}, we get
\[-\frac{d}{dt}\left( 2n\varphi_n+n\frac{Q(t)}{2\pi r}s_n\right)  =\frac{Q(t)}{2\pi r^2}\frac{dr}{dt}n^2 s_n+\frac{n}{\varrho}\varphi_n,\]
or after simplifying
\[2\frac{d\varphi_n}{dt}+ \frac{Q(t)}{2\pi r}\frac{ds_n}{dt}=-\frac{Q(t)}{2\pi r^2}\frac{dr}{dt}(n-1) s_n-\frac{I(t)}{2\pi r}s_n-\frac{1}{\varrho}\varphi_n.\]
Finally, using \eqref{r0} and \eqref{eqSn}
\[\begin{split}
2\frac{d\varphi_n}{dt}+\frac{Q(t)}{2\pi r}\Biggl[-\frac{n}{r}\varphi_n+\varepsilon^\frac{1}{2}\frac{2\pi r\sqrt{\alpha_0}\,\mathrm{sign}(Q(t))}{|Q(t)|}\biggl(1+\frac{2\pi r}{|Q(t)|}\biggr)\frac{n}{r}\varphi_n-\\-\frac{Q(t)}{2\pi r^2}(n-1)s_n-\frac{\varepsilon}{r^2}(n^2-1)s_n-\varepsilon^\frac{1}{2}\frac{(n-1)\sqrt{\alpha_0}}{r}\biggl(1+\frac{2\pi r}{|Q(t)|}\biggr)s_n\Biggr]=\\=-\frac{Q(t)}{2\pi r^2}\biggl(-\frac{Q(t)}{2\pi r}-\frac{\varepsilon}{r}+2\varepsilon^{\frac{1}{2}}\sqrt{\alpha_0}\biggr)(n-1) s_n -\frac{I(t)}{2\pi r}s_n-\frac{1}{\varrho}\varphi_n
\end{split},\]
and after rearranging the terms
\begin{equation}
\begin{split}
\frac{d\varphi_n}{dt}& =\frac{1}{2}\Biggl[\frac{Q(t)}{2\pi r^2}n-\varepsilon^\frac{1}{2}\frac{n\sqrt{\alpha_0}}{r}\left(1+\frac{2\pi r}{|Q(t)|}\right)-\frac{1}{\varrho}\Biggr]\varphi_n+\\&+\Biggr\{\frac{Q(t)}{2\pi r^2}\Biggl[\frac{Q(t)}{2\pi r}+\frac{(n+2)\varepsilon}{2r}+\varepsilon^{\frac{1}{2}}\sqrt{\alpha_0}\biggl(\frac{\pi r}{|Q(t)|}-\frac{1}{2}\biggr)\Biggr](n-1)-\frac{I(t)}{4\pi r}\Biggr\}s_n. 
\end{split}
\label{eqPhin}
\end{equation} 
\end{widetext}

Thus the time evolution of each particular mode has been obtained and it is governed by \eqref{eqSn} and \eqref{eqPhin}.

\subsection{Special limits}

First we study the limit of ideal conductivity. It corresponds to  $\varrho \to 0$, and hence, from \eqref{eqPhin}, we can conclude that $\varphi_n \to 0$. Physically this means that in the limit of very high conductivity, the electric field inside goes to zero ($E_\nu^- \to 0$), as we approach to the behavior of a perfect conductor. If we consider that $Q(t)=Q_0$ is constant or its variation in time is small compared with the evolution of the modes (which also implies $I(t)\to 0$), and the same for the radius of the front $r(t)=r_0$, we can try a solution $s_n = \exp(\omega_n t),\, \varphi_n =0$, to \eqref{eqSn}, and get a discrete dispersion relation of the form
\begin{widetext}
\begin{equation}
\omega_n=-\frac{Q_0}{2\pi r_0^2}(n-1)-\frac{\varepsilon}{r_0^2}(n^2-1)-\varepsilon^\frac{1}{2}\frac{(n-1)\sqrt{\alpha_0}}{r_0}\left(1+\frac{2\pi r_0}{|Q_0|}\right).
\label{disp1}
\end{equation}
\end{widetext}


Next we consider the limit of finite resistivity, but such that the total charge is constant at the surface, or varies very slowly. Writing \eqref{eqPhin} as
\[\frac{d\varphi_n}{dt}=-\frac{d}{dt}\left(\frac{Q(t)}{4\pi r}s_n\right)-\frac{Q(t)}{4\pi r^2}\frac{dr}{dt}n s_n-\frac{1}{2\varrho}\varphi_n,\]
we have now
\begin{equation}
\frac{d\varphi_n}{dt}=-\frac{Q_0}{4\pi r_0}\frac{ds_n}{dt}-\frac{1}{2\varrho}\varphi_n.
\label{dvar}
\end{equation}
For a small enough conductivity, $\varrho \to \infty $ so no extra charge reaches the surface, we find $\varphi_n =-\frac{Q_0}{4\pi r_0}s_n$, and with $s_n = \exp(\omega_n t)$, \eqref{eqSn} yields
\begin{widetext}
\begin{equation}
\omega_n=-\frac{Q_0}{2\pi r_0^2}\left(\frac{n}{2}-1\right)-\frac{\varepsilon}{r_0^2}(n^2-1)-\frac{\varepsilon^\frac{1}{2}\sqrt{\alpha_0}}{r_0}\left(1+\frac{2\pi r_0}{|Q_0|}\right)\left(\frac{3n}{2}-1\right).
\label{disp2}
\end{equation}
\end{widetext}

In a curved geometry we can see that the modes are discrete. However, if we compare \eqref{disp1} and \eqref{disp2}, for small $n$ and vanishingly small $\alpha_0$ there is a $1/2$ factor discrepancy in the dispersion curve between the two limits. The origin of this prefactor was discussed for planar fronts in \cite{CPRL}, and the dispersion relation for planar fronts was obtained in the case of constant charge in \cite{abft}. We get in this 2-D curved case the same factor $1/2$ that we got for the planar case. On the other hand, imposing constant potential at the surface gives a factor of $1$. The intermediate situations can be studied by solving the system \eqref{eqSn} and \eqref{eqPhin}.

Another important consequence is that in both cases the maximum growth correspond to a perturbation with 
\begin{equation}
n \propto|Q_0|/D,
\label{disp3}
\end{equation}
provided that the $\varepsilon^\frac{1}{2}$ term can be neglected and $Q_0$ is negative, implying that the number of fingers increases with the net charge and decreases with electron diffusion.

\subsection{Numerical simulations}
In order to test the analytical predictions, we have calculated numerically the dispersion relation curves for the cases studied previously, when $\varrho \to 0$, so we have a perfect conducting plasma, and when $\varrho$ remains finite.  
We will outline the numerical algorithms and present here the results. 

We start for the case of finite resistivity. The 2-D solution for the potential problem can be written as
\[
\Phi(\mathbf{x})=\int_{\partial\Omega}\frac{1}{2\pi }\log \left\vert \mathbf{x}-\mathbf{x}^{\prime}\right\vert \sigma (\mathbf{x}^{\prime })ds',
\]%
Note that the integration domain $\partial\Omega$ is the curve manifold. The electric field results
\[\mathbf{E}=-\int_{\partial \Omega}\frac{1}{2\pi }\frac{\mathbf{x}-\mathbf{x}^{\prime}}{\left\vert \mathbf{x}-\mathbf{x}^{\prime}\right\vert^2} \sigma (\mathbf{x}^{\prime})ds^{\prime}.
\]
In order to obtain the component in the normal direction $E_{\nu}$, we will multiply it by the normal pointing outside the plasma region, i.e.
\[\mathbf{n}=\frac{(y_{\beta},-x_{\beta})}{\sqrt{x_{\beta}^{2}+ y_{\beta}^{2}}},\]
where the subindex denotes the derivative respect to the curve parameter $\beta$. So we can write
\begin{widetext}
\[E_{\nu} = -\int_{\partial\Omega}\frac{1}{2\pi }\frac{(x-x^{\prime},y-y^{\prime})}{(x-x^{\prime})^{2}+(y-y^{\prime})^{2}} \frac{(y_{\beta},-x_{\beta})}{\sqrt{x_{\beta}^{2}+ y_{\beta}^{2}}}\sigma (x^{\prime},y^{\prime})\sqrt{x^{\prime\, 2}_{\beta}+ y^{\prime\, 2}_{\beta}}d{\beta}^{\prime}.  
\]%
\end{widetext}
Now when approximating the integral as a discrete sum on the interface, i.e. $E_{\nu}^{+}$ limit, some care must be taken. We need the limit $E_{\nu}^{+}$ on the interface. When $\mathbf{x}$ coincides with $\mathbf{x}^{\prime}$ there is an extra contribution of half the pole, which is $\sigma(\mathbf{x})/2$. The $E_{\nu}^{-}$ can be obtained from the boundary condition \eqref{jump}, and the curvature must be expressed in the appropriate coordinates system. 

The case of constant potential, which corresponds to $\varrho=0$, is treated numerically as follows. We have to fulfill the condition
\[
\int_{\partial\Omega}\frac{1}{2\pi }\log \left\vert \mathbf{x}-\mathbf{x}^{\prime
}\right\vert \sigma (\mathbf{x}^{\prime })ds=V_{0},
\]%
being $V_{0}$ a constant for any $\mathbf{x}$ belonging to $\partial\Omega$. Discretizing the domain in small segments $A_{i}$ between points $\mathbf{x}_{i}$ and $\mathbf{x}_{i+1}$ we can approximate the integral as 
\begin{widetext}
\[
\sigma (\overline{\mathbf{x}}_{i})\int_{A_{i}}\frac{1}{2\pi }\log \left\vert 
\overline{\mathbf{x}}_{i}-\mathbf{x}^{\prime }\right\vert ds+\sum_{\substack{
j \\ i\neq j}}\frac{1}{2\pi }\log \left\vert \overline{\mathbf{x}}_{i}-%
\overline{\mathbf{x}}_{j}\right\vert \sigma (\overline{\mathbf{x}}%
_{j})\left\vert \mathbf{x}_{j+1}-\mathbf{x}_{j}\right\vert =V_{0},
\]
where $\overline{\mathbf{x}}_{i}$ is the mean point of the $A_{i}$ segment. The self contribution of the segment to the integral is taken as
\[
\int_{A_{i}}\frac{1}{2\pi }\log \left\vert \overline{\mathbf{x}}_{i}-\mathbf{%
x}^{\prime }\right\vert ds=\int_{-\frac{h_{i}}{2}}^{\frac{h_{i}}{2}}%
\frac{1}{2\pi }\log \left\vert x\right\vert dx=\frac{1}{2\pi }h_{i}\left(
\log \frac{h_{i}}{2}-1\right), 
\]%
\end{widetext}
being $h_{i}$ the length of $A_{i}$. So we end with the equation
\[
M_{ij}\sigma _{j}=V_{0}\mathbf{1},
\]%
where $\mathbf{1}$ is the identity matrix, $\sigma _{j}=\sigma (\overline{\mathbf{x}}_{j})$, and%
\[
M_{ij}=\left\{ 
\begin{array}{c}
\frac{1}{2\pi }h_{j}\log \left\vert \overline{\mathbf{x}}_{i}-\overline{\mathbf{x}%
}_{j}\right\vert,\ \ \ \text{for} \ i\neq j \\ 
\frac{1}{2\pi }h_{i}\left( \log \frac{h_{i}}{2}-1\right), \ \text{for} \ i=j%
\end{array}%
\right. 
\]%
Due to the linearity of the problem, we can solve $A_{ij}\sigma _{j}=\mathbf{%
1}$ and rescale subsequently the solution in order to fulfill $\sum \sigma_{j}h_{j}=Q$.
    
\begin{figure}
\centering
\includegraphics[width=0.5\textwidth]{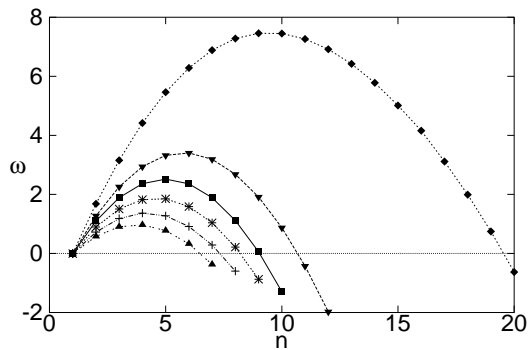}
\caption{Dispersion relation for the discrete modes of a perturbation with initial value $Q=-10$ for different inverse resistivities $1/\varrho$. The  $\blacktriangle$ are for 0, $+$ for 5, $\ast$ for 10, $\blacksquare$ for 15, $\blacktriangledown$ for 25. The case of zero resistivity corresponds to $\blacklozenge$.}
\label{fig3}
\end{figure}

In the numerical simulations presented here, we will follow the evolution of a total initial dimensionless charge $Q=-10$ distributed uniformly along the curve given by 
\begin{eqnarray}
x(\theta) &=& [1+ 0.05\cos(n\theta)]\cos(\theta),\nonumber\\
y(\theta) &=& [1+ 0.05\cos(n\theta)]\sin(\theta).
\label{icharge}
\end{eqnarray}
where n gives the mode of the perturbation and $\theta$ is the curve parameter. We assume that there is not input current, so $j_\nu^{-}=0$ in \eqref{sigmaeq}, and then compute the exponential growth of each mode for a small period of time in order to get the dispersion curve. In Fig.~\ref{fig3}, for different values of the inverse of the resistivity coefficient $1/\varrho$ (or effective conductivity), we plot the corresponding dispersion curves. 

Note that the slope increases with the increase of the conductivity of the plasma, the maxima moves to higher modes, and for larger $n$'s the dispersion curves become negative as predicted by \eqref{disp1} and \eqref{disp2}. The slope around the origin $n=1$ is larger for the case of ideal conductivity, i.e. when the interface is equipotential. 

\section{Comparisons with 2-D positive discharge experiment}

In this section we will make some estimations in order to test the validity of the assumptions made in our contour dynamical model. We will use the experimental data presented at reference \cite{Japos}. The experiment reported there consists in the measuring of the potential and electric field distribution of a surface streamer discharge on a dielectric material. For that, a technique based on Pockels crystals have been applied in order to obtain some temporal and spatial resolution of the discharge (see the reference for details). However, a note of warning must be done: a surface discharge is not a truly 2-D discharge, due to the fact that there is a vertical contribution of the electric field, and the discharge has two different interfaces, the air and the substrate, so the boundary conditions are not the same that the presented so far in this paper. Nevertheless, and keeping that in mind, we may try a quantitative estimation \`a la Fermi from our results and compare it with the actual experiment.   

Here it is a brief account of the experiment. A discharge is created on a dielectric surface using a positive tip and branching is observed. Then the potential is measure using Pockels crystals, laser pulses and a ccd camera. The temporal resolution is 3.2 ns and the electric field close to the tip reaches values of 3 kV/mm, leaving behind a potential gradient of 0.5 kV/mm. At position $r=8$ mm the front moves with an estimated velocity from the pictures of 0.18 mm/ns (from the charge density data, the front has a radius of 4 mm at 3 ns, 8 mm at 15 ns and 9 mm at 28 ns). The pictures show a sharp interface for the charge distribution, so our model should be able to give some quantitative predictions. Unfortunately, there is only one discharge reported, so the estimations we are going to make are very rough.    


The experimental data gives a characteristic front speed $U_0\approx 0.1$ mm/ns, and $\mbox{E}_0 \approx 200$ \,kV/m. In order to get an estimation of the diffusion coefficient $D$ we can make use of the expression \eqref{bb1}. We take \[\mbox{E}_{\nu }^+\approx\frac{3\, \mathrm{kV/mm}}{\mbox{E}_0} \approx 1.5,\,\,\mathrm{and}\,\,\, v_{N}\approx\frac{0.185\,\mathrm{mm/ns}}{U_0} \approx 1.9,\] so that $D\approx 0.05$ is the number that we get. Note that from the expression $U_0=\mu_e \mbox{E}_0$, we could find the experimantal value for the mobility $\mu_e$ for this discharge. 

Now we can make a prediction. The maximum of the dispersion relation will tell us the number of fingers one may find in such experiments. We have calculated the dispersion relation for two limit cases. The limit of ideal conductivity \eqref{disp1} and the limit of infinite resistivity \eqref{disp2}. Those limits would give a lower and upper estimation values for the actual dispersion relation. We expect that the experiment will lie in between and be closer to the predictions given by the limit of infinite resistivity, as the discharge is on a dielectric plate. But before using those dispersion relations we need a further estimation for the surface charge density. We can get the surface density from the jump of the electric field across the interface. So the dimensionless expression reads \[\sigma_0 \approx \frac{(3-0.5)\,\mathrm{kV/mm}}{\mbox{E}_0}, \,\,\,\mathrm{at}\,\, r_0=8 \mathrm{mm}.\] In the dispersion relation expressions \eqref{disp1} and \eqref{disp2} we have to make the substitution $Q_0/2\pi r_0 =\sigma_0$ and find the maximum for $n$. For the ideal conductivity case \eqref{disp1} yields a maximum at $n \approx 76$, and for the case of infinite resistivity \eqref{disp2}, turns out $n \approx 14$. Counting the numbers of real fingers in the experimental pictures at 15 ns, the number is around 20 (one has to extrapolate the number as the pictures do not show the whole discharge). This number is much closer to the lower limit as we pointed before, pointing in the direction that the electrons on the dielectric surface, when moving through the plasma, feel a much higher resistivity than in a conductor.  

Although we do not expect to capture the whole physics of the discharge with the contour model, some essentials ingredients for the early development of the front seem to be well accounted by it. The theoretical prediction made in this section is a rather good one, despite all the approximations made and gives some insight about the parameters involved, such the mobility of the carriers, diffusion coefficient, number of fingers, and so no.  

\section{Conclusions}

We have presented the complete derivation of the contour dynamics moder electric discharges introduced in \cite{Arrayas10}. The model appears as the leading asymptotic description for the minimal streamer model when the electron diffusion coefficient is very small. It consists of two integro-differential equations defined at the boundary of the plasma region: one for the motion of the points of the boundary where the velocity in the normal direction is given in terms of the electric field created by the net charge there, and a second equation for the evolution of the charge density at the boundary. This second equation is very similar to the Taylor-Melcher model in electrohydrodynamics \cite{S}. In the model the electric field is determined by solving Poisson equation with a given surface charge density, leading to a singular integral of the density. 

Once our model has been deduced, we have computed explicit solutions with cylindrical symmetry and investigated their stabilities. The resulting dispersion relation is such as the perturbation with the small mode number can grow exponentially fast. In fact, both the number of modes become unstable and the mode that becomes most unstable (the one corresponding to the dispersion relation) depends critically on the electric resistivity of the media. We have computed analytically the dispersion relation and found that the number of unstable modes grows with the inverse of the resistivity (the conductivity) and the most unstable mode also increases with it. In the limit of vanishing resistivity one can consider the medium as a perfect conductor and therefore impose that the potential is constant at the boundary. The dispersion relation for the model with finite resistivity converges to this limit when resistivity tends to zero.    

We have implemented a numerical procedure to solve our model in general situations. In order to develop the numerical method, we needed to evaluate certain singular integrals that appear when computing the electrical field. As one result of the numerical method, the dispersion relations have been computed and compared them with the analytical results. As a difference to our previous communication \cite{Arrayas10}, we have paid special attention to the effects of Townsend expression for impact ionization \eqref{townsend} on the dispersion relation and the cases of intermediate resistivities. 

Finally, we have taken some experimental data from a positive surface streamer discharge and compare them with our model predictions. The number of fingers calculated from our model is of the same order of the observed one in the actual experiment. We have been able also to estimate the diffusion coefficient from the data. We have shown that the behaviour of the carriers inside the plasma is closer to the limit of high resistivity, so the importance of taking into account the plasma resistivity is made clear. Thus, it is proved that our contour model is able to capture essential parts of the physics involved in the earlier development the streamer discharge, with an extra bonus: we can study more complex geometries and general situations both analytically and numerically. We are now in the process to complete the fully 3-D case and extend these results. 

The authors thank support from the Spanish Ministerio de
Educaci\'on y Ciencia under projects AYA2009-14027-C05-04 and MTM2008-0325.


\begin{thebibliography}{99}

\bibitem{Lagarkov} A. N. Lagarkov and I. M. Rutkevich, {\it Ionization waves in elec-tric breakdown on gases} (Springer-Verlag, New York, 1994).

\bibitem{r1} I. M. Rutkevich, Sov. J. Plasma Phys. {\bf 15}, 844 (1989).

\bibitem{vanS} U. Ebert, W. van Saarloos, and C. Caroli, Phys. Rev. Lett.
{\bf 77}, 4178 (1996); Phys. Rev. E \textbf{55}, 1530 (1997).



\bibitem{aft} M. Array\'as, M. A. Fontelos, and J. L. Trueba, Phys. Rev. E
\textbf{71}, 037401 (2005); J. Phys. A \textbf{39}, 7561 (2006).

\bibitem{r3} A. S. Kyuregyan, Phys. Rev. Lett. \textbf{101}, 174505 (2008)

\bibitem{mmt} M. Array\'as, M. A. Fontelos and J.L. Trueba. J. Phys. D: Appl. Phys \textbf{39} 5176-5182 (2006)

\bibitem{ME} M. Array\'as, U. Ebert, W. Hundsdorfer, Phys. Rev. Lett. \textbf{88}, 174502 (2002).

\bibitem{aftprl} M. Array\'as, M. A. Fontelos, and J. L. Trueba, Phys. Rev. Lett. {\bf 95}, 165001 (2005).

\bibitem{abft} M. Array\'as, S. Betel\'u, M. A. Fontelos, and J. L. Trueba, SIAM J. Appl. Math. {\bf 68}, 1122 (2008).

\bibitem{Dhali} S. K. Dhali and P. F. Williams, Phys. Rev. A {\bf 31}, 1219 (1985); J. Appl. Phys. {\bf 62}, 4696 (1987).

 \bibitem{Vitello} P. A. Vitello, B. M. Penetrante, and J. N. Bardsley, Phys. Rev. E 49, 5574 (1994).

\bibitem{Arrayas10} M. Array\'as, M.A. Fontelos, C. Jim\'enez, Phys. Rev. E {\bf 81}, 035401(R) (2010).

\bibitem{S} D. A. Saville, Annu. Rev. Fluid Mech. {\bf 29} 27--64 (1997).

\bibitem{LF} E.D. Lozansky and O.B. Firsov, J. Phys. D: Appl. Phys. {\bf 6}, 976--981 (1973).
 
\bibitem{contphys} M. Array\'as, J. L. Trueba, Cont. Phys. \textbf{46} 265--276.(2005).

\bibitem{Rai} Y. P. Raizer, {\it Gas Discharge Physics} (Springer, Berlin
1991).

\bibitem{PK} P. Ya. Polubarinova-Kochina, Dokl. Akad Nauk USSR {\bf 47}, no 4, 254.257 (1945) (in Russian).

\bibitem{Low} S. Li, J. S. Lowengrub, J. Fontana, P. Palffy-Muhoray, Phys. Rev. Lett. {\bf 102}, 174501 (2009).

\bibitem{Pismen} L.M. Pismen, {\em Patterns and Interfaces in Dissipative Dynamics}, Springer, where this expansion is explained.

\bibitem{CPRL} M. Array\'as, M. A. Fontelos and J.L. Trueba. Phys. Rev. Lett. {\bf 101}, 139502 (2008).

\bibitem{Japos} D. Tanaka, S. Matsuoka, A. Kumada and K. Hidaka, J. Phys. D: Appl. Phys. \textbf{42}, 075204 (2009).

\end{thebibliography}
\end{document}